%% file: main.tex
\documentclass[acmsmall]{acmart}

\usepackage{algorithmic}
\usepackage{graphicx}
\usepackage{textcomp}
\usepackage{xcolor}
\usepackage{xspace}
\usepackage{balance}
\usepackage{threeparttable}
\usepackage{bbding}
\usepackage{pifont}
\usepackage[normalem]{ulem}
\usepackage{listings}
\usepackage{subcaption}

\input{setups/minted-setup}
\input{setups/tcolorbox-setup}

\AtBeginDocument{%
  }

\newcommand{\eg}{\textit{e.g.},\xspace}
\newcommand{\ie}{\textit{i.e.},\xspace}
\newcommand{\etal}{\textit{et al.}\xspace}
\newcommand{\tool}{WALL-E\xspace}
\newcommand{\base}{\texttt{python-wasmedge}\xspace}

\newcommand{\add}[1]{#1}
\newcommand{\delete}[1]{\ignorespaces}  

\setcopyright{cc}
\setcctype{by}
\acmDOI{10.1145/3808182}
\acmYear{2026}
\acmJournal{PACMSE}
\acmVolume{3}
\acmNumber{FSE}
\acmArticle{FSE175}
\acmMonth{7}
\acmSubmissionID{fse26mainb-p1581-p}
\received{2025-09-11}
\received[accepted]{2026-03-24}

\begin{document}

\title{Bringing Managed Language Support to WebAssembly with External Library Linking}

\author{Shuyao Jiang}
\orcid{0000-0002-2797-875X}
\affiliation{%
  \department{Department of Computer Science and Engineering}
  \institution{The Chinese University of Hong Kong}
  \city{Hong Kong}
  \country{China}
}
\email{syjiang21@cse.cuhk.edu.hk}

\author{Ruiying Zeng}
\orcid{0009-0000-6367-0077}
\affiliation{%
  \department{College of Computer Science and Artificial Intelligence}
  \institution{Fudan University}
  \city{Shanghai}
  \country{China}
}
\email{ryzeng22@m.fudan.edu.cn}

\author{Yangfan Zhou}
\orcid{0000-0002-9184-7383}
\authornote{Yangfan Zhou is the corresponding author.}
\affiliation{%
  \department{College of Computer Science and Artificial Intelligence}
  \institution{Fudan University}
  \city{Shanghai}
  \country{China}
}
\email{zyf@fudan.edu.cn}

\author{Michael R. Lyu}
\orcid{0000-0002-3666-5798}
\affiliation{%
  \department{Department of Computer Science and Engineering}
  \institution{The Chinese University of Hong Kong}
  \city{Hong Kong}
  \country{China}
}
\email{lyu@cse.cuhk.edu.hk}


\begin{abstract}
  WebAssembly (Wasm) has emerged as a powerful bytecode format for running applications with near-native performance in portable and secure environments. However, while Wasm currently supports compiled languages like C, C++, and Rust, it lacks robust support for managed languages such as Python, Java, and JavaScript. This limitation hinders the deployment of applications in domains like machine learning and data processing that rely heavily on managed language ecosystems. 
  To address this, we propose \tool, a novel framework to integrate managed languages into Wasm environments without complex runtime nesting or recompilation. \tool employs a unique external library linking strategy, using a client-server architecture to connect Wasm modules with managed language libraries running in their native runtimes. This approach preserves the native execution speed and \delete{full feature support} \add{language feature compatibility} of managed languages by eliminating the overhead associated with double-layer virtual machines. \delete{The framework also includes an LLM-assisted programming interface to assist cross-language development.}
  Our evaluation shows that \tool supports ten managed languages without framework modifications and achieves a speedup of hundreds of times over the runtime nesting solution, with low communication overhead.
  \tool enhances the practicality of Wasm in cloud and edge computing, enabling efficient multi-language applications.
\end{abstract}



\maketitle

\input{tex/1-intro}

\input{tex/2-bg}

\input{tex/3-app}
\input{tex/4-eval}
\input{tex/5-dis}

\input{tex/6-rw}
\input{tex/7-con}

\begin{acks}
This work was supported by the National Natural Science Foundation of China (Project No. 62572127), the Research Grants Council of the Hong Kong Special Administrative Region, China (No. CUHK 14209124 of the General Research Fund), and RGC Grant for Theme-based Research Scheme Project (RGC Ref. No. T43-513/23-N).
\end{acks}

\bibliographystyle{ACM-Reference-Format}
\bibliography{reference}


\end{document}

%% file: setups/minted-setup.tex
\usepackage[frozencache=true,cachedir=minted-cache]{minted}

\definecolor{codebg}{RGB}{248,248,248}
\definecolor{framecolor}{RGB}{200,200,200}

\setminted{
    fontfamily=tt,
    fontsize=\scriptsize,
    breaklines=true,
    breakautoindent=true,
    linenos=true,
    numberblanklines=true,
    numbersep=8pt,
    frame=single,
    framesep=10pt,
    rulecolor=framecolor,
    framerule=0.8pt,
    bgcolor=codebg,
    autogobble=true,
    tabsize=4,
    obeytabs=false,
    showspaces=false,
    showtabs=false,
    escapeinside=||,
    mathescape=false,
    firstnumber=auto,
    stepnumber=1,
    baselinestretch=1.0,
    xleftmargin=15pt,
    xrightmargin=15pt,
}

\setminted[python]{
    style=default,     
    numbersep=10pt      
}

\setminted[java]{
    style=vs,
    linenos=true
}

\setminted[javascript]{
    style=monokai,
    numbersep=6pt
}

\setminted[json]{
    style=paraiso-light,
    linenos=true
}

\setminted[cpp]{
    style=borland,
    frame=lines
}

\setminted[html]{
    style=friendly,
    linenos=false
}

\newminted{python}{}
\newminted{java}{}
\newminted{javascript}{}
\newminted{json}{}
\newminted{cpp}{}
\newminted{html}{}

\setmintedinline{
    fontsize=\small,
    bgcolor=codebg
}

%% file: setups/tcolorbox-setup.tex
\usepackage{tcolorbox}
\tcbuselibrary{skins, breakable} 

\newtcolorbox{mybox}{
    colback=gray!5!white,    
    colframe=gray!75!black,  
    arc=3pt,                 
    boxrule=1pt,             
    breakable,               
    enhanced                  
}

%% file: tex/1-intro.tex
\section{Introduction}\label{sec:intro}

WebAssembly (abbreviated Wasm)~\cite{haas2017bringing} is a binary instruction format designed as a compilation target for high-level programming languages.
It was originally used for computationally intensive web applications and has been supported by all major browsers~\cite{wagner2017webassembly}.
Wasm provides near-native speed, a memory-safe execution environment, cross-platform portability, and lightweight bytecode size.
Such advantages also make Wasm increasingly popular outside the web.
In recent years, Wasm has been widely adopted on many server-side applications, \eg cloud computing~\cite{shillaker2020faasm,gackstatter2022pushing,gadepalli2020sledge}, smart contracts~\cite{zhang2024vm,wang2020wana,chen2022wasai}, and microcontrollers~\cite{gurdeep2019warduino,zandberg2021femto}.

As Wasm gains popularity in various fields, a crucial issue for its future development is the \textbf{capability of programming language support}, \ie enabling various programming languages to integrate with Wasm.
Wasm is used as a compilation target, meaning that Wasm programs are not written by hand but compiled from high-level source languages.
Therefore, to promote Wasm across more scenarios, it is essential to provide comprehensive language support for Wasm.
However, the current language support for Wasm is still immature.
The source languages that Wasm fully supports are limited, including some typical compiled languages, \ie C, C++, and Rust~\cite{hilbig2021empirical}.
But many other languages, especially \textit{managed languages} (\eg Python, Java, and JavaScript), still lack an effective mechanism for interacting with Wasm.
As a result, a large number of existing applications written in different managed languages (\eg machine learning applications in Python, database applications in Java) cannot be effectively deployed via Wasm, which is a critical bottleneck in the development of the Wasm ecosystem.

The key challenge in bringing managed language support to Wasm is that the execution of such languages relies on their own \textit{managed runtimes}, \eg the Python interpreter.
\delete{While} \add{At the same time,} Wasm programs also need to run on Wasm runtimes, \eg WasmEdge~\cite{wasmedge}.
Thus, to support the execution of a managed language in the Wasm context, it is inevitable for the Wasm runtime to handle the specific managed runtime of this source language (called \textit{external runtime}).
Unfortunately, there is still no satisfactory solution for Wasm to handle those external runtimes.
The current mainstream mechanism is \textit{runtime nesting}, which first compiles the external runtime (typically written in compiled languages such as C/C++) to Wasm bytecode and then runs it on the Wasm runtime, followed by executing the source program on that Wasm-formatted runtime~\cite{wasmlabs-python}.
This solution is intuitive but has several non-ignorable disadvantages.
First, it is \textit{non-extensible} for different source languages. Compiling the external runtime is a complex task that requires domain knowledge of the source language to ensure the correctness of the runtime functionality. So, it is hard to maintain the runtime compilation for different source languages and even different versions of the same source language.
Second, it leads to \textit{poor performance}, \ie the program executes at a slow speed. This is obvious since runtime nesting introduces dual virtual environments, which increase system-level overhead during runtime, such as context switching and resource management.
Third, the Wasm-formatted runtime can only support \textit{limited language features}. 
For example, \base~\cite{wasmlabs-python}, a state-of-the-art solution that converts CPython~\cite{cpython} (a typical Python interpreter) to Wasm bytecode, still lacks support for many important Python packages.

To address the challenge of external runtime handling, we propose a novel framework \textbf{\tool} (\textbf{WA}sm \textbf{L}anguage \textbf{L}inker via \textbf{E}xternal Library) for bringing managed language support to Wasm.
The key idea is \textbf{external library linking}: Take the target application written in a managed language (\textit{external language}, \eg Python) as an external library, which keeps the application running on its original runtime.
Then, use a startup program written in a language natively supported by Wasm (\textit{native language}, \eg Rust) to link the external library dynamically. 
\delete{The premise of this design is that the external library is trustworthy, so the safety feature provided by Wasm can be guaranteed.}
\add{The premise of this design is that external libraries execute in a trusted environment. The Wasm module remains sandboxed, while the external execution is controlled by the trusted host runtime.}
Compared to the runtime nesting solution, our design has the following advantages:

\begin{itemize}
    \item \textbf{\add{Language} extensibility:} \tool does not require compiling the external runtimes into the Wasm format, so it is extensible to \delete{any} \add{most} external language without complicated maintenance.
    \item \textbf{High performance:} \tool keeps external libraries running on their original runtimes, which can achieve a \delete{much} faster execution speed than the runtime nesting solution since it avoids much system-level overhead (\eg context switch).
    \item \textbf{\delete{Complete support} \add{Language feature compatibility}:} Without compiling the external runtimes into the Wasm format, \delete{\tool can provide complete support for the external languages (\ie support full language features of the external languages) based on their original runtimes} \add{\tool can leverage the original runtime ecosystems and thus support a broad set of language features of external languages}.
\end{itemize}

To implement \tool, we need to consider two critical issues.
The first one is how to design the architecture of \tool to support the extensibility of programming languages. The second one is how to provide good usability, \ie make it easy for users to use \tool.
To achieve language extensibility, \tool adopts HTTP communication for external library linking since it is a universal communication protocol supported by any programming language.
Specifically, \tool provides a startup program written in Rust as the HTTP client, and the target external library is encapsulated as the HTTP server. The client program is compiled into Wasm bytecode and runs on the Wasm runtime, and the server program runs on its original runtime as a web service.
When starting library invocation, the client sends an HTTP request (including the data required by the target library) to the server to call the external library. Then, the server handles the request and returns the execution result of the target library to the client.
To achieve good usability, \tool provides a unified interface to users for external library invocation. \tool also provides automatic server-side deployment, so users need only focus on client-side invocation logic.
\delete{Furthermore, \tool incorporates an LLM-assisted programming interface that provides intelligent cross-language development support.}

We conducted comprehensive experiments to evaluate the effectiveness of \tool.
First, to evaluate the language extensibility, we apply \tool to link different types of external libraries written in ten popular managed languages. We showed that \tool is extensible across different managed languages and flexible to use in various applications.
Second, to evaluate the performance of \tool, we compared the execution speed of \tool and the runtime nesting solution. We made the testing application run on the original runtime (our solution) and the Wasm-formatted runtime (the runtime nesting solution), then measured the execution speed of the application under both scenarios. The results indicated that \tool achieved hundreds of times faster execution speed than the runtime nesting solution.
Third, we further measured the communication overhead of \tool and found that it accounts for only a small portion of the total process time.

In summary, this work makes the following contributions:
\begin{itemize}
    \item \textbf{Novel Concept:} We present the first systematic study on managed language support for Wasm and introduce the innovative approach of external library linking to enable efficient integration of managed languages with Wasm.
    \item \textbf{Framework Design:} We design and implement \tool, a practical framework that achieves managed language support in Wasm with multi-language extensibility, high performance, and \delete{full} \add{broad language} feature compatibility.
    \item \textbf{Comprehensive Evaluation:} We apply \tool across multiple managed languages and conduct extensive experiments, demonstrating its effectiveness in real-world scenarios.
\end{itemize}

%% file: tex/2-bg.tex
\section{Background}\label{sec:bg}

\subsection{Wasm and Its Applications}

WebAssembly (Wasm)~\cite{haas2017bringing} is a binary instruction format designed as a portable compilation target for high-level programming languages, enabling efficient execution on the web and beyond. Initially developed to complement JavaScript in browsers, Wasm provides near-native performance by leveraging a stack-based virtual machine optimized for speed, security, and compact bytecode~\cite{reiser2017accelerate}. Its sandboxed execution model ensures memory safety and platform independence, making it suitable for diverse environments.

In browser-based applications, Wasm excels in performance-critical tasks such as game engines, multimedia processing, and scientific simulations, where JavaScript falls short in efficiency. For instance, frameworks like Unity~\cite{unity} and Unreal Engine~\cite{unreal} compile to Wasm to deliver high-fidelity graphics in web applications. 
Beyond the browser, the advantages of Wasm have spurred its adoption in server-side environments, including cloud computing~\cite{jain2022extending}, edge computing~\cite{gackstatter2022pushing}, and serverless architectures~\cite{shillaker2020faasm}. 
Projects like Wasmtime~\cite{wasmtime} and WasmEdge~\cite{wasmedge} extend Wasm's capabilities to standalone runtime environments, enabling developers to run Wasm modules on servers with low overhead and cross-platform compatibility.

Given Wasm's expanding role in both browser and server-side environments, 
\add{its ability to provide high-performance support for multiple programming languages is crucial.}
\delete{its ability to serve as a universal compilation target for multiple programming languages is crucial.
A wide range of applications, from high-performance web applications to cloud-native workloads, can benefit from the portability and efficiency of Wasm, but only if developers can leverage their preferred programming languages. 
Supporting diverse languages, including those with different memory models and runtime requirements, would further broaden the adoption of Wasm, enabling seamless integration of existing codebases and fostering a more inclusive ecosystem.}
\add{This need becomes evident in modern platforms that execute third-party extensions inside a Wasm sandbox on the critical request path. A representative example is checkout-time customization logic (\eg discounts, shipping, and payment rules) deployed as Wasm modules in e-commerce platforms. The platforms often enforce a strict latency budget (on the order of a few milliseconds) to avoid slowing down user-facing transactions~\cite{shopify}.
In such settings, developers often wish to reuse mature managed language ecosystems (\eg Python/JavaScript libraries for business rules, data processing, or lightweight analytics). However, supporting managed languages within the Wasm execution model remains non-trivial in practice. This motivates mechanisms for integrating managed languages into the Wasm ecosystem under tight latency and deployment constraints.}

\subsection{Language Support Mechanism of Wasm}

Despite its growing ecosystem, the effectiveness of Wasm still hinges on robust support for diverse programming languages. While \textit{compiled languages} like C/C++ and Rust have mature Wasm toolchains, many \textit{managed languages}, those with garbage collection or dynamic typing (\eg Python, Java), face challenges due to the linear memory model of Wasm and lack of built-in runtime features. 
We then discuss the language support mechanism of Wasm and how it motivates our work.

\begin{figure}[t]
  \centering
  \begin{subfigure}[t]{\textwidth}
    \centering
    \includegraphics[width=0.6\linewidth]{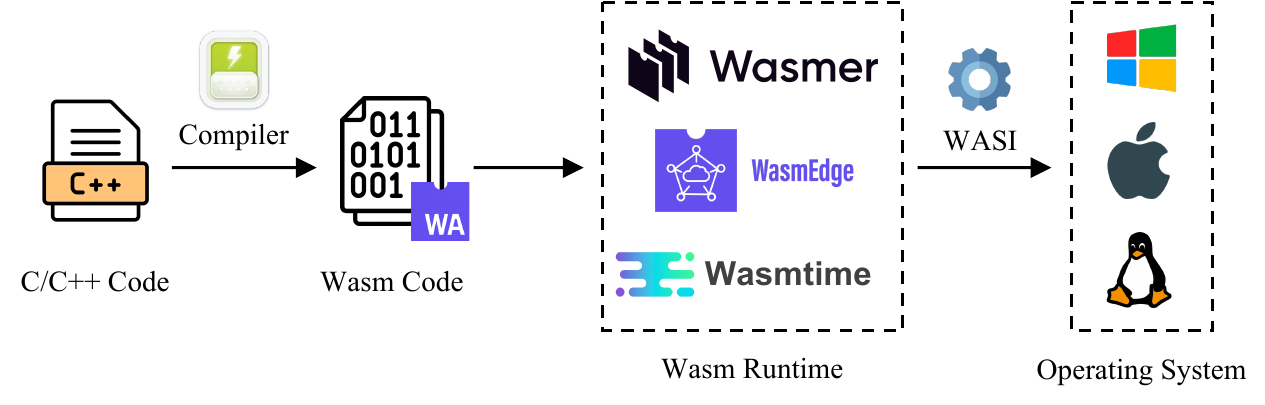}
    \vspace{-5pt}
    \caption{\footnotesize{\textbf{Compiled Language:} Workflow with C/C++ as the source language}}
    \label{fig:bg_a}
  \end{subfigure}

  \vspace{\baselineskip}
  
  \begin{subfigure}[t]{\textwidth}
    \centering
    \includegraphics[width=0.75\linewidth]{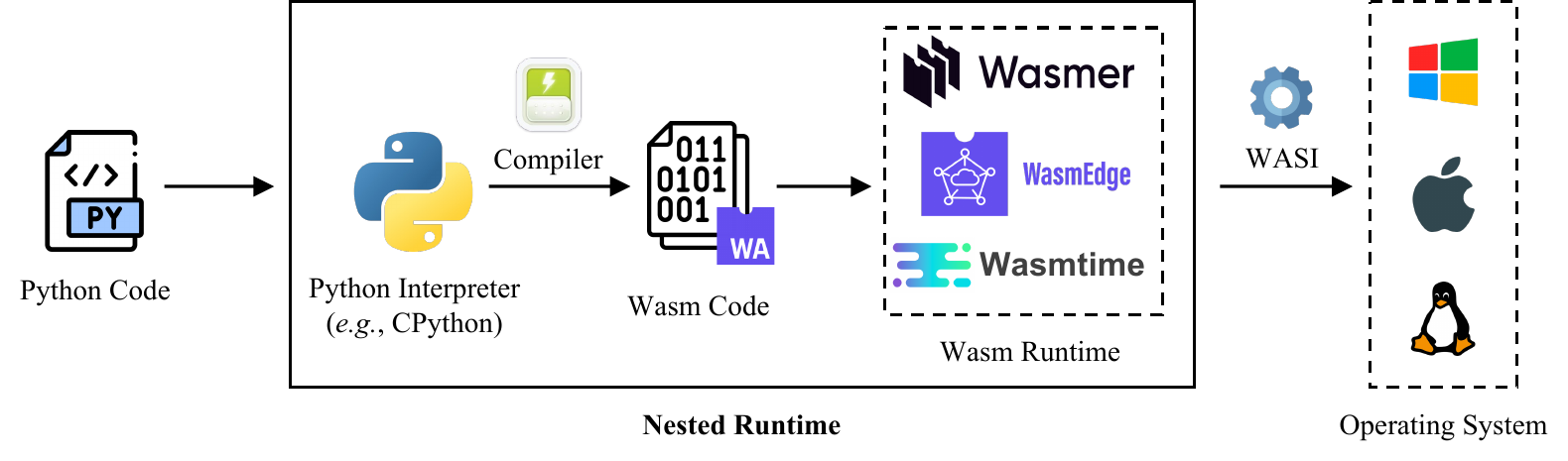}
    \vspace{-5pt}
    \caption{\footnotesize{\textbf{Managed Language:} Workflow with Python as the source language}}
    \label{fig:bg_b}
  \end{subfigure}
  
  \caption{Current language support mechanism of Wasm for different types of programming languages.}
  \label{fig:background}
  
  \vspace{-5pt}
\end{figure}


\subsubsection*{\textbf{Compiled Language Support}}
The execution of programs written in compiled languages (\eg C, C++, Rust) to Wasm follows a well-defined toolchain-based workflow. Source code is first transformed into Wasm bytecode through language-specific compilers that target the Wasm Intermediate Representation (Wasm IR). A prominent example is Emscripten~\cite{zakai2011emscripten}, an LLVM-based toolchain that compiles C/C++ source code to optimized Wasm modules while generating necessary JavaScript glue code for browser integration. This glue code serves as a bridge between the compiled Wasm module and the browser's JavaScript engine (\eg V8~\cite{v8}, SpiderMonkey~\cite{spider-monkey}), handling memory management and facilitating system calls through standardized Web APIs.

For non-browser environments, the WebAssembly System Interface (WASI)~\cite{clark2019standardizing} specification provides a portable, capability-based security model for system interactions. WASI enables compiled Wasm modules to safely access operating system resources (file I/O, network sockets, etc.) while maintaining Wasm's core security guarantees. Standalone Wasm runtimes (\eg Wasmtime~\cite{wasmtime}, Wasmer~\cite{wasmer}, and WasmEdge~\cite{wasmedge}) implement WASI, executing Wasm modules in isolated sandboxes with near-native performance. 
The typical workflow is as shown in Figure~\ref{fig:bg_a}.
These runtimes employ advanced optimization techniques, including ahead-of-time (AOT) compilation and tiered just-in-time (JIT) compilation, to maximize execution efficiency.
This robust support for compiled languages has made Wasm an attractive compilation target for performance-critical applications ranging from multimedia processing to scientific computing, while maintaining cross-platform compatibility and security.

\subsubsection*{\textbf{Managed Language Support}}
The design of Wasm as a low-level compilation target presents inherent challenges when executing programs written in managed languages (\eg Python, JavaScript, and Java), due to their reliance on high-level runtime environments. 
Unlike compiled languages that can be directly translated into Wasm bytecode, managed languages require their runtime systems to execute properly. 
Current solutions attempt to address this through \textbf{runtime nesting}, a technique where the language’s runtime (typically implemented in a systems language like C) is first compiled into Wasm, allowing the source program to run on top of this Wasm-based runtime. 
Figure~\ref{fig:bg_b} shows the typical workflow with Python as the source language.

Unfortunately, this solution introduces several fundamental limitations that constrain its practical application.
First, \textit{the implementation is non-extensible}. Porting a language runtime to Wasm requires deep expertise in both the source language implementation and the execution model of Wasm. Each new language or even version update demands significant engineering effort to recompile and validate the runtime, making the solution difficult to maintain and scale across different managed languages. 
Second, \textit{the performance is suboptimal}, as the code must undergo multiple layers of interpretation, from the original language's bytecode to Wasm bytecode and finally to machine code. This multi-level interpretation introduces substantial overhead. 
Third, \textit{the language feature support is limited}, as many advanced features that rely on direct system interactions or native extensions cannot function properly within Wasm's constrained execution environment.

These limitations manifest clearly in specific language implementations.
For Python, while CPython has been compiled to Wasm for executing Python code, the runtime nesting solution makes it difficult to support different Python versions or alternative implementations. Performance suffers due to the double interpretation layer, and critical Python packages are unsupported, as seen in the limited functionality of \base~\cite{wasmlabs-python}.
JavaScript support similarly demonstrates these weaknesses. For example, WasmEdge's integration of QuickJS~\cite{quickjs} provides only basic JavaScript functionality and exhibits poor performance compared to native V8 execution.
The challenges are most pronounced for Java, where no practical Wasm solution exists. The JVM's complexity, with its JIT compilation, sophisticated garbage collection, and extensive native interfaces, makes it particularly difficult to adapt through runtime nesting. 

The inherent limitations of runtime nesting, including poor extensibility, performance overhead, and limited feature support, reveal the need for a fundamentally different approach to running managed languages on Wasm. Therefore, we attempt to design a novel mechanism that eliminates runtime nesting while achieving \delete{full} \add{broad} language support and native performance.

%% file: tex/3-app.tex
\section{\tool: Managed Language Support Framework for Wasm}\label{sec:app}

\begin{figure}[t]
    \centering
    \includegraphics[width=0.95\linewidth]{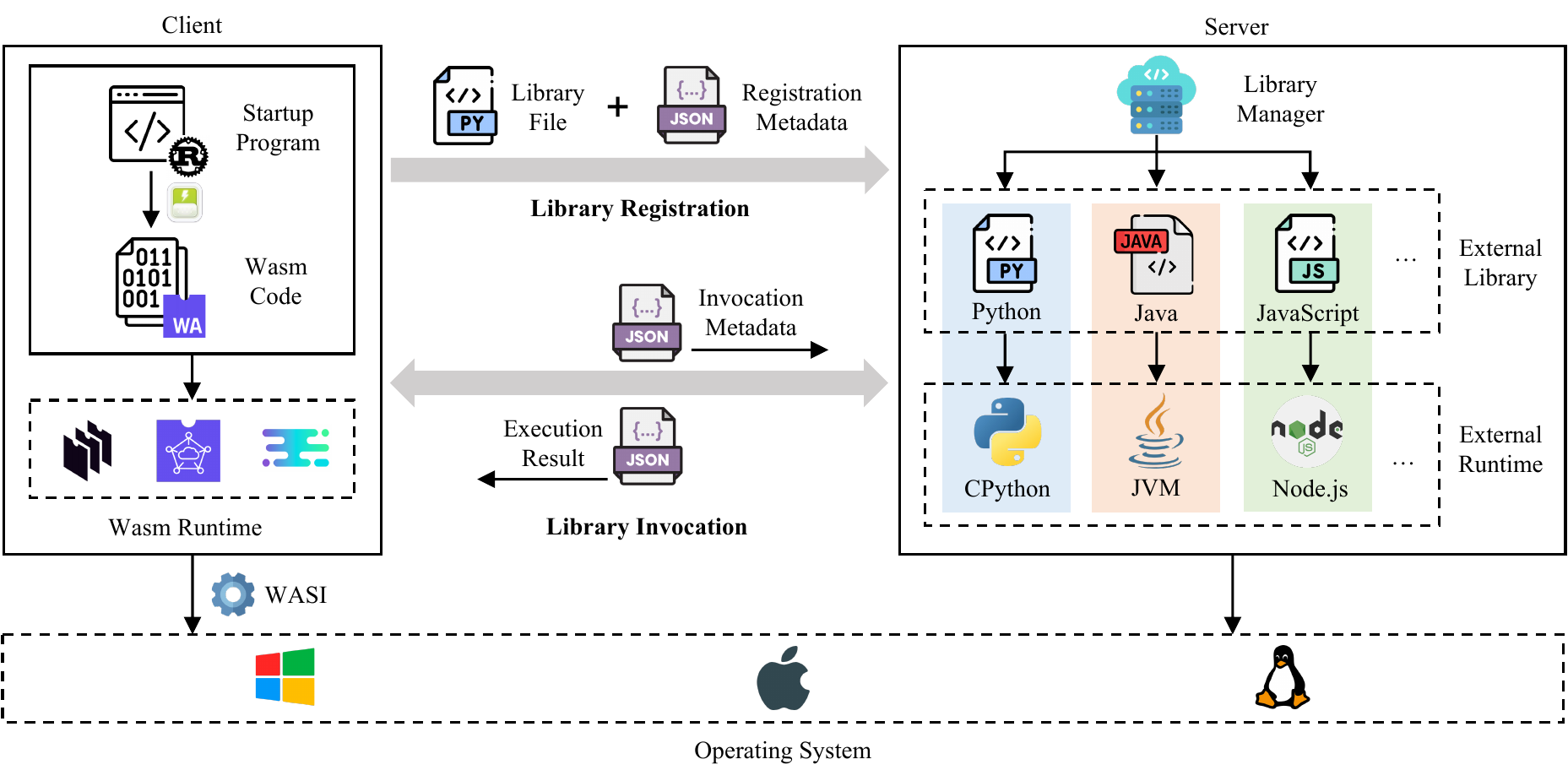}
    \caption{\add{The framework design of \tool: A client-server architecture that registers and invokes external libraries via HTTP while executing library code in native runtimes.}}
    \label{fig:overview}
    \vspace{-5pt}
\end{figure}

To address the fundamental limitations of runtime nesting described above, we propose \textbf{\tool}, a \delete{brand-new} framework for supporting managed languages for Wasm, based on the core idea of \textbf{external library linking}. 
\tool treats managed language applications (written in \textit{external language} such as Python) as \delete{pre-}prepared external libraries that remain executing on their own runtimes, while a lightweight Wasm-native startup program (written in \textit{native language} such as Rust) dynamically \delete{links and coordinates} \add{binds to and invokes} these external libraries. 
\delete{This design operates under the trust boundary that external libraries are vetted, preserving Wasm's security guarantees.}
\delete{Compared to conventional runtime nesting, \tool achieves three transformative advantages:
(1) \textit{Extensibility}. By eliminating the need to recompile language runtimes to Wasm, our solution supports any managed language without complex maintenance;
(2) \textit{High performance}. Retaining execution on native runtimes avoids system-level overhead like context switching, delivering near-native speed; 
(3) \textit{Complete support}. Full language feature support is maintained since all functionality executes through the original, unmodified runtime environment. This architectural innovation fundamentally changes how Wasm integrates with managed languages.}
\add{Our design assumes that external libraries execute in trusted environments. The Wasm-side execution remains sandboxed, while the security properties of external execution rely on the host runtime and deployment controls.}
\add{Compared to conventional runtime nesting, \tool is designed to improve: 
(1) \textit{Language extensibility}, by avoiding recompilation of language runtimes into Wasm; 
(2) \textit{Execution performance}, by eliminating nested-runtime interpretation and executing managed code in mature runtimes; and 
(3) \textit{Language feature compatibility}, by leveraging the original runtime ecosystems without modifying the Wasm execution model.}

Figure~\ref{fig:overview} shows the overall framework design of \tool.
To achieve language extensibility, the \tool framework implements a \textit{client-server architecture} through standardized HTTP-based communication, chosen for its universal support across programming ecosystems. 
The workflow of \tool operates through multiple coordinated components.
First, during the library registration phase (\textbf{Section~\ref{sec:app_1}}), the client user submits external library files accompanied by structured metadata descriptors, while the server infrastructure dynamically allocates network endpoints and instantiates managed service instances through a centralized library manager. 
The core library invocation phase (\textbf{Section~\ref{sec:app_2}}) \delete{employs a sophisticated parameter marshaling system where client-specified invocation metadata undergoes automated transformation to adapt to the external language before execution.}
\add{uses a metadata-driven parameter marshaling mechanism: The client encodes invocation requests with explicit type annotations, and the external runtime reconstructs native values based on the declared types.}
Execution results are serialized and returned via HTTP channels, completing the cross-runtime invocation cycle. 
\delete{To streamline developer interaction, \tool incorporates an LLM-assisted programming interface (\textbf{Section 3.3}) capable of optimizing parameter conversions, dependency management, and error handling.} 
\add{The implementation of \tool is based on the HTTP services provided by WasmEdge (\textbf{Section~\ref{sec:impl}}).}
Overall, the architectural design of \tool ensures broad language extensibility through protocol standardization, \add{and} maintains native execution performance by preserving original runtime contexts\delete{, and enhances usability via AI-powered developer tooling}.
Next, we will elaborate on the design \add{and implementation} details of \tool.

\subsection{Library Registration}\label{sec:app_1}

Before calling an external library, the user first needs to conduct library registration, \ie provide the library files with \add{a} structured metadata \delete{descriptors} to the library manager.
\add{Here, metadata refers to a JSON descriptor that specifies the library name, language, service endpoint, available function signatures (\eg function name, parameters, and return type), and optional dependencies, which the library manager uses for service instantiation and invocation routing.}
\delete{The library registration phase serves as the gatekeeper in \tool's architecture, addressing two critical requirements in cross-language Wasm ecosystems: \textit{security assurance} and \textit{operational usability}.
First, from a security perspective, registration establishes a mandatory trust boundary between the trusted Wasm sandbox and potentially untrusted external libraries. While Wasm modules operate within a secure sandboxed environment, external libraries contain user-defined code that executes natively on their respective runtimes, posing potential security risks to the host system. Mandatory registration enables systematic security validation before any cross-runtime invocation occurs. 
Second, registration significantly enhances usability and manageability for developers. By decoupling library deployment from invocation logic, \tool eliminates the need for manual infrastructure configuration. Users simply submit their library files and metadata through a standardized interface, and the system automatically handles service deployment, endpoint allocation, and health monitoring. This abstraction becomes particularly valuable in large-scale deployment scenarios where dozens of external libraries may be active simultaneously.}
\add{The registration phase provides a centralized mechanism for validating and configuring external libraries before they can be invoked from Wasm. 
From a security standpoint, registration requires external libraries to be explicitly declared and added to an allowlist, helping prevent accidental invocation of undeclared endpoints. While Wasm modules remain sandboxed, external libraries execute in their native runtimes, and their security properties depend on the host environment and deployment controls rather than the Wasm sandbox itself. Accordingly, registration focuses on validating metadata and performing basic sanity checks prior to deployment, but it does not provide sandboxing for the external runtimes, which are assumed to be trusted in our deployment model.
From the usability perspective, registration also decouples library deployment from invocation by automating service instantiation and endpoint allocation based on the submitted metadata. This separation reduces client-side configuration effort and provides a consistent interface for managing multiple external libraries.}
For a clearer description, we introduce the registration workflow from the perspectives of the client and the server, respectively.

\subsubsection*{\textbf{Client-Side Submission Process}}
The registration workflow begins with the client-side preparation of the external library.
To simplify the user operation process, \tool provides a standardized startup program for end users. For library registration, the user only needs to provide two essential components to the library manager:

\begin{itemize}
    \item \textbf{Library Source Code:} The actual implementation files (\eg Python modules, Java classes) containing the functional logic
    \item \textbf{Structured Metadata:} A JSON descriptor specifying the operational characteristics of the provided library
\end{itemize}

We choose JSON as the metadata description format due to its exceptional cross-language compatibility and universal support across programming ecosystems. This selection ensures that library metadata can be generated, parsed, and processed by both the client and server.
For user convenience, \tool provides a template of the structured metadata. Based on the template, the user needs to declare the following fields of the library: \texttt{"name"}, \texttt{"language"}, \texttt{"endpoint"}, \texttt{"functions"}, and \texttt{"dependencies"}. In particular, the \texttt{"endpoint"} field refers to the IP address of the external library as a web service, which can be auto-assigned by \tool or defined by the user. In the \texttt{"functions"} field, the user can declare multiple related functions, each of which should specify the \texttt{"name"}, \texttt{"params"}, and \texttt{"return\_type"}.
Figure~\ref{fig:reg_metadata} shows an example registration metadata of a Python library named \texttt{image-processor}, providing various functional modules for image processing. In this example, we set the value of the \texttt{"endpoint"} field as \texttt{"auto"}, indicating that \tool will automatically assign an IP address to this library.

\input{code/meta_reg}

The client then sends a simple HTTP POST request to the \texttt{/register} endpoint of the library manager, using multipart form data to transmit both the code files and metadata simultaneously. This approach eliminates complex client-side configuration, requiring only minimal information from developers.

\subsubsection*{\textbf{Server-Side Automated Deployment}}
Upon receiving a registration request, \tool initiates \delete{a comprehensive} \add{an} automated deployment pipeline that transforms the client-submitted data into a \delete{fully operational} \add{runnable} service. 
This process begins with a \textit{validation phase} where the library manager \delete{systematically} verifies the metadata structure, ensuring all required fields (\eg service name, language specification, and function definitions) are present and properly formatted. 
\delete{The library manager then performs security compliance checks on the library code, scanning for prohibited operations such as unrestricted filesystem access or unauthorized network calls, thereby maintaining the security integrity of the overall Wasm ecosystem.}
\add{The library manager then performs basic sanity checks on the library code to flag disallowed access patterns. 
Concretely, the checks focus on validating the submitted package (\eg rejecting path traversal patterns and invalid file layouts) and scanning for a limited set of filesystem/network access patterns inconsistent with the declared library configuration (\eg attempts to access unavailable filesystem resources or initiate outbound network calls).
These checks are intended to reduce obvious misconfigurations and unsafe patterns, rather than to provide sandboxing for external runtimes.}

Following successful validation, \tool proceeds to \textit{resource allocation}, where it automatically assigns necessary network and computational resources. A unique service identifier is generated using UUID-based naming conventions (\eg \texttt{lib-python-a1b2c3}), providing a distinct reference for the new library service. The deployment engine dynamically allocates available IP addresses and ports \delete{from the managed resource pool, ensuring collision-free network endpoint assignment} \add{for the service to avoid endpoint conflicts}. Based on the specified language requirements, the system can also configure computational resources (\eg memory allocation and CPU limits) \delete{to ensure optimal runtime performance} \add{when supported by the deployment environment}.

When the resource is ready, \tool conducts \textit{service instantiation} that executes language-specific initialization procedures. For example, for Python libraries, the deployment engine automatically creates a Flask~\cite{flask} server scaffold with pre-configured endpoints including \texttt{/invoke} for function execution and \texttt{/health} for \delete{service monitoring} \add{liveness checks}. Similarly, for Node.js applications, the system configures an Express~\cite{express} server with appropriate middleware and routing structures. 
\delete{Each service is containerized with security constraints and resource limits, providing isolation while maintaining the original runtime environment required for proper library execution.}
\add{In our implementation, each service can be deployed in an isolated environment (\eg a container) to separate external runtimes, while preserving the original runtime required for library execution.}

The deployment pipeline concludes with \textit{health check}, where \tool monitors the newly instantiated service until it becomes \delete{fully operational} \add{responsive}. Through periodic HTTP requests to the health check endpoint, the library manager verifies service responsiveness. This process ensures that only properly initialized and responsive services are registered in the discovery system, preventing incomplete deployments from affecting the overall system reliability. Upon a successful health check, the service is marked as active and becomes available for client-side invocations.

\subsection{Cross-Runtime Library Invocation}\label{sec:app_2}

The library invocation process in \tool establishes a \delete{seamless} bridge between Wasm modules and external language runtimes through a \delete{well-designed} client-server interaction model. Similar to the library registration workflow, the invocation process also follows a clear division between client-side request preparation and server-side execution management.
A critical challenge in this phase is achieving \textit{cross-language parameter sharing}, as different programming languages employ fundamentally different type systems and data representation formats that must be reconciled for successful interoperation. Thus, we design a \delete{sophisticated} parameter marshaling system for \tool to achieve automated parameter transformation between the client and the server.

\subsubsection*{\textbf{Client-Side Invocation Preparation}}
The invocation workflow begins with the client application preparing a structured invocation request containing all necessary information for cross-runtime execution. Since the target library has been registered as required, the information that the user needs to provide for invocation can be further simplified.
The client constructs a JSON payload as the invocation metadata that specifies:

\begin{itemize}
    \item \textbf{Target Identification:} The unique service identifier (\texttt{"library\_id"}) obtained during library registration
    \item \textbf{Function Specification:} The exact function name (\texttt{"func"}) and parameter list (\texttt{"params"}) for execution
\end{itemize}

The request format maintains consistency with the registration metadata schema, creating a unified interface across both processes.
Figure~\ref{fig:invo_metadata} provides an example of a typical invocation request targeting the Python library \texttt{image-processor} described in Section~\ref{sec:app_1}. In this example, the user aims to invoke the function \texttt{enhance\_contrast} in the target library, with two parameters of type ``ndarray'' and ``float''.

\input{code/meta_invo}

This example demonstrates how \tool encapsulates the external language parameters from the client side. To enable seamless data exchange between disparate language runtimes, we employ a language-neutral parameter encoding strategy where all parameter values are represented as strings with explicit type annotations.
For basic parameter types such as integers, floats, and strings, we utilize a straightforward string representation combined with an explicit value. A float value of 1.5, for instance, is encoded as \texttt{\{"type": "float", "value": "1.5"\}}, ensuring that both strongly-typed Wasm modules and dynamically-typed Python runtimes can reconstruct the original value.
\delete{For complex parameter types like NumPy arrays, we implement a more sophisticated encoding scheme that preserves both data content and structural metadata. Multi-dimensional arrays are serialized using base64 encoding for binary data preservation, accompanied by explicit shape and data type information.}
\add{For complex parameter types that cannot be directly represented as JSON primitives (string/number/boolean), we use base64 encoding to preserve their byte-level representation. This mainly applies to raw binary payloads (\eg image buffers or serialized objects) and numerical arrays (\eg NumPy arrays). For example, for NumPy arrays, we encode the raw byte buffer using base64 and attach structural information, including the explicit shape and data type.}
A typical array parameter might be encoded as \texttt{\{"type": "ndarray", "value": "\delete{base64\_encoded\_data} \add{BASE64\_ENCODED\_DATA}", "shape": [256, 256], "dtype": "float32"\}}\add{, where \texttt{"BASE64\_ENCODED\_DATA"} represents the actual base64-encoded string of the raw bytes}.
This approach enables the reconstruction of exact array structures on the server side, maintaining data integrity across the language boundary. 

The client subsequently transmits this structured JSON payload to the server's \texttt{/invoke} endpoint via HTTP POST, initiating the cross-runtime library execution pipeline.


\subsubsection*{\textbf{Server-Side Request Processing}}

Upon receiving an invocation request, the server executes a \delete{sophisticated} multi-stage processing pipeline that ensures \delete{secure and efficient} cross-runtime execution. This structured approach transforms client-submitted requests into actual function executions within the target language runtime \delete{while maintaining safety and efficiency} \add{with consistent request handling}.

The processing pipeline begins with \textit{service discovery}, where the library manager resolves the provided \texttt{library\_id} to the actual service endpoint using the centralized service registry. The library manager queries the database of the registered libraries to retrieve the complete service metadata, including the network endpoint, supported functions, and runtime characteristics. This service resolution mechanism provides \delete{a critical abstraction layer that enables dynamic load balancing and fault handling without client-side awareness}
\add{an abstraction layer that enables dynamic service binding without client-side configuration}.

After successful service resolution, the library manager performs \textit{request forwarding} where the original invocation request (including unprocessed parameters in their JSON format) is transparently sent to the target external runtime. The manager constructs an HTTP POST request to the \texttt{/invoke} endpoint of the external service, preserving all original parameters and metadata. This design ensures that each language runtime maintains full control over parameter parsing and validation, leveraging its native type system for accurate data conversion.

The core execution occurs through \textit{distributed parameter processing} where each external runtime independently handles parameter parsing and function invocation. Upon receiving the forwarded request, the embedded web server of the target runtime (\eg Flask for Python, Express for Node.js) parses the JSON parameters using its native type system.
Take Python as an example, this involves converting JSON representations into actual Python objects through the \add{built-in} \texttt{parse\_parameter()} function, which handles base64-decoding of NumPy arrays, string-to-number conversions, and other language-specific transformations.
\add{For user-defined Python classes, \texttt{parse\_parameter()} attempts to resolve the declared type against class definitions in the registered library and reconstructs the object accordingly. If resolution fails or a mismatch occurs, the service returns an error message.
This resolution relies on the consistency between the submitted library files and the type declarations in the metadata, but it does not require users to implement custom (un)marshaling logic on the client side.}
\add{If parameter parsing succeeds,} the runtime then executes the requested function with the properly typed parameters and captures the execution results.

Finally, the process completes with \textit{result response} where the external runtime returns the execution results to the library manager, which then forwards the final response to the client. The external runtime serializes the function return values using its native \texttt{convert\_result()} mechanism, ensuring complex data types like NumPy arrays are properly encoded as base64 strings with appropriate metadata. 
The library manager acts as a transparent proxy, adding minimal overhead while providing \delete{essential services such as request logging, performance monitoring, and error handling normalization} \add{request forwarding, optional logging, and a consistent response format} across different language runtimes.


\noindent\delete{\textbf{3.3 LLM-assisted Programming}}


\delete{
Although the design of \tool has minimized client-side operations to the greatest extent, we recognize that developers may still face cognitive barriers when working with cross-language invocations. To provide an optimal user experience, \tool introduces a Large Language Model (LLM) Programming Interface that significantly enhances developer productivity in cross-language Wasm environments. 
This integration addresses the fundamental challenge of bridging disparate programming ecosystems by reducing the cognitive overhead associated with understanding multiple type systems, serialization formats, and error handling patterns. Empirical evidence from recent research~\cite{nam2024using,yang2024cref,denny2024explaining} demonstrates that LLMs have remarkable capabilities in comprehending and generating code across multiple programming languages, making them ideally suited for assisting with cross-language development tasks. 
The LLM-assisted features in \tool specifically target three critical areas where developers typically encounter the greatest challenges: cross-language parameter mapping, dependency management, and error handling, thereby accelerating the usage of diverse language libraries.
}

\delete{
\textbf{Cross-language Parameter Mapping}.
From the user's perspective, a critical challenge in cross-language external library invocation is parameter representation and transmission. To address this, \tool provides an LLM-assisted parameter mapping functionality that transforms diverse language data types into standardized JSON representations for HTTP transmission. 
Developers interact with the LLM interface using a structured prompt template that specifies essential context, including target library information, parameter mapping request, and response requirements. 
The template ensures the LLM receives complete information to generate tailored JSON serialization strategies, including exact parameter structures, encoding instructions for binary data, validation rules, and error handling guidance. 
We present an example prompt template for parameter mapping below, and due to space limitations, other templates are left in the supplementary material.
}

\delete{
Based on this structured template, the user just needs to input the target language and function signatures for the library information. The user can specify the parameter mapping request, \eg "How to structure the JSON parameters for calling this function?" For additional information, the user can describe the use case for this invocation, \eg "calling Python image processing libraries in microservices". The user can also add other supplementary information.
}

\delete{
\textbf{Dependency Management}.
Another critical issue in cross-library invocation involves identifying and managing the required dependencies and runtime configurations for external libraries. The LLM interface provides intelligent dependency management by analyzing library metadata and function signatures to generate comprehensive dependency specifications. When the user provides information about the target external library, the LLM identifies necessary packages, version constraints, and environment configurations specific to the library's language ecosystem.
For example, the LLM can suggest appropriate pip packages and version ranges for Python libraries, and npm packages with compatibility information for Node.js libraries. 
Additionally, it offers dependency resolution strategies for handling version conflicts and cross-language compatibility issues, along with validation rules to ensure the dependency configuration will function correctly within the Wasm environment.
}

\delete{
\textbf{Error Handling}.
The LLM interface in \tool also provides intelligent error diagnosis and repair for cross-library invocation issues. When developers encounter errors, the LLM analyzes failure patterns across parameter serialization, type mismatches, dependency conflicts, and runtime exceptions. 
For parameter errors, it identifies issues like incorrect base64 encoding or missing metadata, providing corrected JSON examples and implementation guidance. For dependency problems, it suggests version adjustments and configuration changes. 
With sophisticated prompts, the LLM can generate specific remediation strategies with step-by-step correction procedures and preventive best practices.
This LLM-assisted error handling mechanism transforms error messages into actionable solutions, significantly reducing debugging time and improving development efficiency within the cross-language environment.
}

\subsection{Implementation}\label{sec:impl}

\delete{\textbf{Wasm Runtime Framework}.}
The implementation of \tool leverages WasmEdge's HTTP services~\cite{wasmedge-http} to provide the Wasm runtime environment for client-side operations, with Rust as the source language of the client-side startup program. \add{Our implementation is built on WasmEdge 0.14.0 and Rust toolchain 1.27.1.}
We select WasmEdge as the runtime framework for its excellent support for HTTP services, lightweight footprint, and high-performance execution, making it ideal for edge computing scenarios. 
The implementation uses WasmEdge's async-enabled HTTP handler system to create an invocation gateway that receives JSON payloads from client applications, routes requests to appropriate external libraries, and serializes responses back to Wasm modules.
\add{We implement the user-submitted JSON metadata as a dynamically loaded configuration file, so updating library bindings only requires modifying the configuration without rebuilding the framework code.}
This runtime foundation ensures sandboxed client-side execution while maintaining low-latency communication with external language services through optimized HTTP protocols.
It is worth noting that \tool is designed as a generic framework that can be implemented using various standalone Wasm runtimes beyond WasmEdge, providing flexibility for different deployment environments.

\delete{
\textbf{LLM Interface}.
For the LLM-assisted programming features, \tool implements a standalone RESTful service that provides provider-agnostic LLM capabilities. The service offers asynchronous support through thread pooling while maintaining compatibility with multiple large language model providers. 
The initial implementation integrates OpenAI's GPT-4 API~\cite{openai}, providing immediate access to advanced code generation capabilities. 
The interface employs carefully engineered prompt templates that transform developer queries into structured requests for type mapping, dependency resolution, and error diagnosis. A caching layer stores frequently requested patterns to reduce API calls, while a validation component ensures generated code meets the security and performance standards of \tool before being presented to developers.
}

%% file: code/meta_reg.tex
\begin{figure}[t]
\centering
\begin{jsoncode}
{
  "name": "image-processor",
  "language": "python",
  "endpoint": "auto",  // Server auto-assigns or user defines
  "functions": [
    {
      "name": "enhance_contrast",
      "params": [
        {"name": "input_image", "type": "ndarray"},
        {"name": "enhancement_factor", "type": "float"}
      ],
      "return_type": "ndarray"
    }
  ],
  "dependencies": ["numpy", "opencv-python"]
}
\end{jsoncode}
\caption{Example registration metadata of a Python library.}
\label{fig:reg_metadata}
\vspace{-5pt}
\end{figure}

%% file: code/meta_invo.tex
\begin{figure}[t]
\centering
\begin{jsoncode}
{
  "library_id": "lib-python-a1b2c3",
  "func": "enhance_contrast",
  "params": [
    {"type": "ndarray", "value": "BASE64_ENCODED_DATA", "shape": [256, 256], "dtype": "float32"},
    {"type": "float", "value": "1.5"}
  ]
}    
\end{jsoncode}
\caption{Example invocation metadata of a Python library.}
\label{fig:invo_metadata}
\vspace{-5pt}
\end{figure}

%% file: tex/4-eval.tex
\section{Evaluation}\label{sec:eval}

In this section, we evaluate the effectiveness of \tool in supporting managed languages within the Wasm environment.
We aim to address the following research questions:

\begin{itemize}
    \item \textbf{RQ1 (Language Extensibility):} How extensible is \tool in supporting various managed languages without complex modifications?
    \item \textbf{RQ2 (Execution Performance):} How does the execution speed of applications using \tool compare to the existing runtime nesting solution?
    \item \textbf{RQ3 (Communication Overhead):} What is the impact of the client-server communication on the performance of \tool?
\end{itemize}

\subsubsection*{\textbf{Experiment Environment}}
All our experiments are conducted on a server with a 2-core Intel(R) Xeon(R) Platinum 8259CL CPU @ 2.50GHz and 16GB RAM. The operating system is 64-bit Ubuntu 24.04 LTS with the Linux kernel version 6.8.0-1009-aws.
The experimental settings related to each RQ will be introduced in the following subsections.

\subsection{RQ1: Language Extensibility}

\subsubsection*{\textbf{Settings}}
Given that language extensibility represents the most fundamental design criterion of \tool, our evaluation specifically targets this core capability through systematic testing across diverse programming languages. 
We selected the top 10 managed languages from \textit{The RedMonk Programming Language Rankings}~\cite{redmonk}, representing diverse programming paradigms and runtime characteristics. 
The tested languages include JavaScript, Python, Java, PHP, C\#, TypeScript, Ruby, R, Scala, and Kotlin. 
Each language was configured with its dominant web server framework\delete{and representative real-world applications from its respective ecosystem}
\add{, and tested on representative benchmarks and real-world applications}
to ensure a comprehensive evaluation of the language support capability of \tool. 
\delete{The evaluation focused on assessing the framework's ability to integrate diverse managed languages without requiring modifications to the core WALL-E infrastructure.}

\add{We first evaluate the language support capability of \tool on language-specific benchmarks using a measurable criterion.
We report a \textit{pass rate} for each language, defined as the fraction of benchmark programs that execute successfully through \tool and produce the expected output.
To avoid bias in benchmark selection, we follow three principles: 
(1) Use public and widely recognized benchmark suites; 
(2) Use functionally similar workloads across languages to reduce language bias; and 
(3) The benchmarks can be executed independently without a complex harness, so that pass/fail outcomes can be directly observed.
Based on these principles, we select benchmark programs from \textit{The Computer Language Benchmarks Game}~\cite{benchmarksgame}, which is widely used for comparing programming language performance across diverse computational tasks and provides implementations in more than twenty languages.}

\add{For each language, we include \textit{all} benchmark programs available for that language in the Benchmarks Game.
We first execute these programs in a native local environment to establish a baseline of runnable tests. 
Programs that pass locally are then deployed as services in \tool using the corresponding language-specific web server, and re-executed through \tool with the same standard inputs. 
This two-stage setup separates failures caused by language toolchain or environment issues from those introduced by \tool integration.
A program is considered \textit{pass} in \tool if it completes without errors and produces the expected output; otherwise, it is marked as \textit{fail} (\eg runtime error or output mismatch). 
We report the pass rate of \tool on locally runnable programs for each language as the primary measure of language support.}

\add{In addition to the benchmark testing, we also configure each language with representative real-world applications as supplementary evidence of practical compatibility. 
These application-level experiments aim to assess \tool's ability to integrate with common frameworks and deployment setups in practice, complementing the benchmark-based evaluation.}

\add{Table~\ref{tab:languages} summarizes the 10 selected managed languages, including their current Wasm support status, runtime versions, representative web server frameworks, and applications used in our experimental setup.
The ``Limited'' in the Wasm support status means that the source language can be supported by runtime nesting with limited language features.
Table~\ref{tab:bm_intro} overviews the computational tasks included in the benchmark suites, where each task has multiple implementation versions on each language.
We next report the benchmark pass rates and application-level integration results of \tool for the evaluated languages.
}

\input{tables/languages}

\input{tables/bm_intro}

\input{tables/bm_pass}

\subsubsection*{\textbf{Results}}

\add{Based on the benchmark suite, \tool achieves a high overall pass rate across the 10 managed languages. As shown in Table~\ref{tab:bm_pass}, out of 278 locally runnable tests, 266 pass when executed via \tool, achieving an overall pass rate of 95.68\%. Eight languages (JavaScript, Java, C\#, TypeScript, Ruby, R, Scala, and Kotlin) achieve a 100\% pass rate on all their runnable tests. The remaining failures are concentrated in Python (27/30, 90\%) and PHP (22/31, 70.97\%).
Overall, these results show that \tool can be empirically evaluated with measurable success criteria and observable negative outcomes, providing quantitative evidence of its language extensibility.
}

\input{figures/python_fail_exp}

\add{We then conduct a detailed case analysis for the failed cases. Figure~\ref{fig:python_fail_exp} shows a representative Python failed case \texttt{spectralnorm\_py\_4}, which raises \texttt{``NameError: pool is not defined''} when invoked as an HTTP endpoint. 
The root cause is that this benchmark relies on a module-level multiprocessing \texttt{Pool} that is initialized only inside the \texttt{if \_\_name\_\_ == "\_\_main\_\_":} block (shown in Figure~\ref{fig:python_fail_bm}). 
When deployed as a long-running web service, the benchmark module is imported rather than executed as a script (shown in Figure~\ref{fig:python_fail_app}), so the \texttt{\_\_main\_\_} initialization is bypassed and the global \texttt{pool} is never created. As a result, calls to \texttt{pool.starmap(...)} fail at runtime. 
The other two failed cases (\texttt{spectralnorm\_py\_7} and \texttt{mandelbrot\_py\_5}) are also caused by a similar reason.
This failure highlights a class of scripts that depend on global initialization, which requires a lightweight adapter (\eg explicitly initializing such global resources in the service wrapper) when exposing them through our invocation interface.}

\input{figures/php_fail_exp}

\add{The nine failed PHP tests are represented by \texttt{binarytrees\_php\_6} (shown in Figure~\ref{fig:php_fail}), whose output differs from the local CLI baseline. This benchmark is implemented as a multi-process program: It uses \texttt{pcntl\_fork()}/\texttt{pcntl\_waitpid()} to spawn worker processes and relies on shared memory primitives to aggregate per-depth results. 
When deployed as a web service, such process control primitives are commonly restricted or disabled for safety and resource isolation, which can alter the program's execution or aggregation behavior and introduce missing lines or runtime warnings in the output. 
As a result, the service response no longer matches the expected CLI output, leading to an output-mismatch failure under our correctness criterion. 
This case highlights that benchmarks that depend on OS-level process management may require adaptation (\eg, a single-process fallback) when exposed through an HTTP invocation interface.}

\add{Beyond the benchmark suite, all selected real-world applications in Table~\ref{tab:languages} were successfully deployed and executed through \tool. These applications cover diverse domains, such as OCR processing (Tesseract.js), enterprise search (Elasticsearch), statistical visualization (ggplot2), and network operations (MSQuic). The results provide additional evidence that \tool can integrate practical workloads from multiple managed language ecosystems in realistic deployments.}

\delete{The experimental results demonstrate the excellent language extensibility of \tool, successfully supporting all ten managed languages without requiring framework modifications. 
Table~\ref{tab:languages} shows the detailed information on the ten tested managed languages, including their runtime versions, current Wasm support status, \tool support status, with the corresponding web server frameworks and applications adopted in \tool.
The "Limited" in the current support status means that the source language can be supported by runtime nesting with limited language features.
As shown in Table~\ref{tab:languages}, \tool achieved perfect compatibility across diverse programming ecosystems. 
This comprehensive support stands in stark contrast to the current limited or non-existent Wasm support for these languages, particularly for JavaScript, Java, TypeScript, R, Scala, and Kotlin, which currently lack viable Wasm compilation pathways.}

\delete{Notably, \tool seamlessly integrated with each language's dominant web server framework, including Express for JavaScript/TypeScript, Flask for Python, Spring Boot for Java, ASP.NET Core for C\#, Sinatra for Ruby, and modern frameworks like Javalin for Kotlin and Scalatra for Scala. 
The successful deployment of representative real-world applications, ranging from Tesseract.js for OCR processing and Elasticsearch for enterprise search to ggplot2 for statistical visualization and MSQuic for networking operations, validates the practical utility of \tool across diverse application domains.}

\delete{The results confirm that the external library linking approach in \tool effectively bypasses the traditional limitations of Wasm language support, providing a unified invocation mechanism that works consistently across different runtime environments and programming paradigms. This demonstrates that language extensibility can be achieved through architectural innovation rather than through complex compiler modifications or runtime adaptations, establishing a new paradigm for multi-language support in Wasm ecosystems.}

\begin{mybox}
    \textbf{Summary of RQ1:} \delete{\tool demonstrates excellent language extensibility by successfully supporting all ten tested managed languages through the external library linking strategy, achieving broad compatibility across diverse programming ecosystems.}
    \add{\tool achieves broad language extensibility, reaching an overall benchmark pass rate of 95.68\% across 10 managed languages and successfully deploying all representative real-world applications.}
\end{mybox}

\subsection{RQ2: Execution Performance}

\subsubsection*{\textbf{Settings}}
Since enhancing the execution performance of managed languages is also a crucial design objective for \tool, we conducted a comprehensive evaluation to compare the runtime performance of various applications between \tool and the existing runtime nesting solution. 
We selected Python as the target language, as it represents a typical managed language with significant runtime overhead. The baseline for comparison is the \base runtime developed by VMware Labs~\cite{py-wasmedge}, an authoritative implementation that provides Python runtime capabilities within Wasm through runtime nesting techniques.
To ensure a fair comparison, we maintained identical software versions (WasmEdge 0.14.0 and Python 3.12) across both environments.
\add{It is worth noting that RQ2 reports the \textit{execution time of benchmark code} measured inside the Python runtime, excluding the client-server communication overhead introduced by \tool. We evaluate such end-to-end overhead separately in RQ3.}

In terms of benchmark selection, we initially attempted to use the standard Python performance test suite \textit{PyPerformance}~\cite{pyperf} to ensure comparability with established practices. However, we encountered a fundamental limitation: The test suite relies on packages that currently lack WASI support and therefore cannot execute within the \base runtime. This constraint prevented the use of standardized benchmark suites and necessitated the creation of a custom benchmark dataset.
To address this challenge, we constructed a  benchmark suite consisting of 15 distinct tests covering three major workload categories: 

\begin{itemize}
    \item \textbf{Compute-intensive Tasks:} Pystone, N-Queens problem solving, Prime number calculation, Matrix multiplication, Fibonacci sequence computation
    \item \textbf{I/O-intensive Tasks:} File I/O, JSON serialization/deserialization, Text processing, List comprehensions, Dictionary operations
    \item \textbf{Memory-intensive Tasks:} Memory allocation, String manipulations, Large data structure creation, Object instantiation, Recursive structure processing
\end{itemize}

\add{In addition to the micro-benchmarks above, we further include two real-world Python libraries to assess \tool on practical software and non-trivial interaction patterns. Specifically, we evaluate JSON schema validation using \texttt{jsonschema}~\cite{jsonschema} and Markdown rendering using \texttt{markdown}~\cite{markdown}.
For \texttt{jsonschema}, we validate a dataset of $n{=}1024$ documents; for \texttt{markdown}, we render $n{=}256$ Markdown inputs. To model non-trivial interactions beyond a single call, we vary the invocation granularity by splitting each workload into $k \in {1, 8, 32}$ chunks and invoking the external runtime $k$ times per run, while keeping the total input size $n$ fixed. 
This setup enables us to study both execution performance under realistic application logic and the impact of repeated cross-runtime invocations under the same workload.}

\add{For each workload (including micro-benchmarks and real-world libraries), we run it 10 times in both environments and report the average execution time, excluding the first run to reduce initialization effects. This setting provides a consistent basis for comparison under the package and WASI constraints of the \base runtime.}

\delete{Each benchmark was carefully designed to exercise specific aspects of the Python runtime while maintaining compatibility with both \tool and \base.
The experimental protocol involved executing each benchmark 10 times in both environments and calculating average execution times (excluding the first execution to eliminate the effect of initialization overhead) to ensure statistical reliability. This approach allowed for meaningful performance comparisons while working within the technical constraints of Wasm runtimes, particularly the limited package availability in WASI-compliant Python distributions.}

\subsubsection*{\textbf{Results}}

\input{tables/exec_time}

\input{tables/exec_py_apps}

Table~\ref{tab:exec_time} presents a detailed comparison of the execution time of the Python benchmarks run on both \base and \tool. For each benchmark, the table records the average execution time and calculates the speedup ($\uparrow\times$) achieved by \tool \delete{, providing a comprehensive overview of its performance across diverse workloads}.
The results reveal that \tool delivers significant performance enhancements across all categories, with the most significant gains in compute-intensive tasks (average 647$\times$ speedup), followed by I/O-intensive (500$\times$) and memory-intensive (386$\times$) operations. 
This performance hierarchy directly reflects the cumulative overhead imposed by the nested runtime architecture of \base. The exceptionally high speedups in computational tasks like ``matrix'' (959$\times$) and ``prime'' (833$\times$) highlight the tremendous cost of double virtualization on pure CPU operations, while the cost is also significant for I/O tasks and memory-intensive operations.

\add{Table~\ref{tab:exec_py_apps} reports the execution time of two real-world Python libraries. \tool consistently outperforms \base by large margins across all chunk settings. For \texttt{jsonschema} ($n{=}1024$), \tool achieves $689\times$ speedup on average. For \texttt{markdown} ($n{=}256$), the average speedup is $449\times$. Notably, varying the chunk number ($k \in {1,8,32}$) does not materially change the execution time within each environment, indicating that the dominant factor is the underlying execution stack rather than the invocation granularity.}

These performance gaps are directly attributable to their fundamental architectural differences. 
The \base runtime uses a nested architecture, \add{incurring substantial overhead from nested execution and cross-boundary transitions.}
\delete{where the Python interpreter runs within a Wasm runtime, leading to severe performance degradation. This design introduces massive overhead from double virtualization and expensive cross-boundary calls between the two runtime layers.
In contrast, the key innovation of \tool is executing the application directly within a native, unified runtime environment.}
\add{In contrast, \tool executes the workload in the native Python runtime instead of embedding the interpreter into Wasm. The execution is triggered via network communication through its client, which invokes the native Python runtime as an external service.}
\add{As a result, \tool avoids the nested-runtime overhead and achieves consistently lower execution times across workloads.}
\delete{This approach fundamentally eliminates the nested virtualization overhead, avoiding the context switching penalties and resource duplication that plague the nested model, thereby drastically reducing execution times across all operation types.}

\begin{mybox}
    \textbf{Summary of RQ2:} \delete{\tool demonstrates significant performance improvements over \base, achieving average speedups of 647$\times$, 500$\times$, and 386$\times$ for compute-intensive, I/O-intensive, and memory-intensive tasks respectively, by eliminating the overhead of nested runtime virtualization through its native execution architecture.}
    \add{\tool substantially outperforms \base by avoiding nested-runtime execution. Across micro-benchmarks, \tool achieves average speedups of 647$\times$, 500$\times$, and 386$\times$ for three workload categories. On real-world libraries, \tool delivers 689$\times$ (\texttt{jsonschema}) and 449$\times$ (\texttt{markdown}) speedups on average.}
\end{mybox}

\subsection{RQ3: Communication Overhead}

\subsubsection*{\textbf{Settings}}

\begin{figure}[t]
    \centering
    \includegraphics[width=\linewidth]{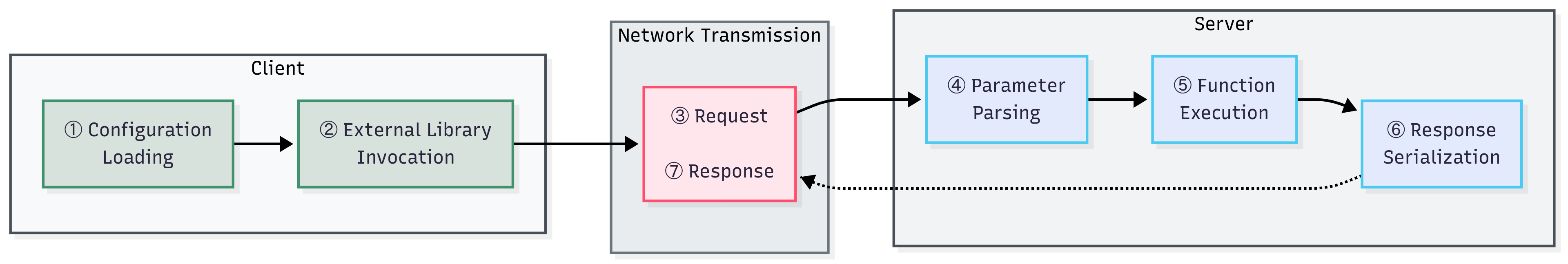}
    \caption{The complete external library invocation process in \tool.}
    \label{fig:inovke_flow}
    \vspace{-5pt}
\end{figure}

The client-server architecture of \tool inherently introduces communication overhead due to the HTTP-based interaction between Wasm modules and external runtimes. 
To quantitatively evaluate the impact of this overhead on overall performance, we conducted a fine-grained time analysis of the complete external library invocation process. 
As illustrated in Figure~\ref{fig:inovke_flow}, the invocation workflow is decomposed into seven distinct phases: (1) configuration loading on the client side, (2) external library invocation, (3) HTTP request transmission, (4) parameter parsing on the server, (5) function execution, (6) response serialization, and (7) HTTP response transmission.

To measure the duration of each phase, we conducted instrumentation within the framework source code. We inserted timestamps before and after each critical operation, ensuring microsecond-level precision. This instrumentation strategy allowed us to capture the exact time consumption of each processing stage without significantly affecting the overall system performance.
A key consideration involves the measurement of network transmission time, which spans both client and server environments. We observed that the client-measured external library invocation period (\ding{193}) actually includes all the subsequent phases (\ding{194}\textasciitilde\ding{198}).
Therefore, rather than attempting synchronized cross-runtime timing, we calculated the network duration (\ding{194}+\ding{198}) by subtracting the total server-side processing time (\ding{195}+\ding{196}+\ding{197}) from the external library invocation time (\ding{193}). This approach eliminates the need for clock synchronization while providing accurate estimates of network latency.

For overhead analysis, we consider all phases except the actual function execution (\ding{196}) as the total communication-related overhead. 
This includes configuration loading, parameter serialization and deserialization, network transmission, and response processing. The proportional contribution of these overhead components to the entire process time (\ding{192}+\ding{193}) serves as the primary metric for evaluating the efficiency of the cross-language communication mechanism of \tool.

\subsubsection*{\textbf{Results}}

\input{tables/overhead}

\input{tables/overhead_py_apps}

Our experimental results reveal a consistent pattern of communication overhead across all tests in \tool. Due to space constraints, we present detailed results only for compute-intensive benchmarks in Table~\ref{tab:overhead}, which show that the total communication overhead remains low, averaging only 0.30\% of the total process time. \delete{This minimal overhead demonstrates the efficiency of the client-server communication mechanism of \tool despite the cross-language invocation requirements.}
\add{The overhead distribution is consistent across different workloads. Network transmission constitutes the dominant overhead component, accounting for 0.26\% on average. Client-side processing overhead remains negligible at 0.02\%, consisting of configuration loading and parameter serialization. Server-side overhead is also minimal at 0.02\%, including parameter parsing and response serialization. The actual function execution time accounts for 99.70\% of the total process time on average, indicating that the vast majority of time is spent on productive computation rather than on framework overhead.}

\delete{The overhead distribution shows consistent characteristics across different workloads. Network transmission constitutes the dominant overhead component, accounting for 0.26\% of total time on average, which includes both request and response transmission phases. Client-side processing overhead remains negligible at 0.02\%, primarily consisting of configuration loading with parameter serialization. Server-side overhead is equally minimal at 0.02\%, encompassing parameter parsing and response serialization operations.
The actual function execution time represents 99.70\% of the total process time on average, indicating that the vast majority of time is spent on productive computation rather than framework overhead. This efficiency is consistent across diverse computational patterns, from mathematical calculations in the "prime" benchmark (84.37\% computation time) to complex recursive computations in the "fibonacci" benchmark (99.88\% computation time).}

\add{Table~\ref{tab:overhead_py_apps} further reports overhead statistics for the real-world libraries. 
On average, the communication overhead accounts for 11.39\% of the total time for \texttt{jsonschema} and 2.25\% for \texttt{markdown}, and the overhead ratios remain stable across $k \in \{1,8,32\}$. 
The overhead for \texttt{jsonschema} is higher because each run involves larger request/response payloads (a schema plus $n{=}1024$ JSON documents), leading to more expensive data serialization, transmission, and JSON parsing than \texttt{markdown} ($n{=}256$ texts). 
Despite this increase, the overhead remains acceptable in practice, especially given the substantial execution-time improvements observed in RQ2.
}

\begin{mybox}
    \textbf{Summary of RQ3:} \delete{The client-server communication of \tool introduces minimal overhead, accounting for only 0.30\% of total process time on average, demonstrating that cross-language invocation costs are negligible compared to actual computation time.}
    \add{\tool incurs low communication overhead overall. It accounts for 0.30\% of total time on micro-benchmarks, and remains modest on real-world workloads (11.39\% for \texttt{jsonschema} and 2.25\% for \texttt{markdown}), indicating that cross-language invocation costs are generally small relative to execution time.}
\end{mybox}

%% file: tables/languages.tex
\begin{table}[t!]\small
\centering
\caption{\add{Managed languages for the language extensibility evaluation.}}
\vspace{-5pt}
\begin{threeparttable}
\add{
\begin{tabular}{l|c|lll}
 \toprule
 \textbf{Language\tnote{*}} & \textbf{Wasm Support} & \textbf{Runtime} & \textbf{Web Server} & \textbf{Application} \\
 \midrule
  JavaScript & \XSolidBrush & Node.js v20.16.0 & Express & Tesseract.js \\
  Python & Limited & CPython 3.12.3 & Flask & Face recognition \\
  Java & \XSolidBrush  & OpenJDK 21.0.8 & Spring Boot & Elasticsearch \\
  PHP  & Limited & PHP 8.3.10 & Slim-server & Image processing \\
  C\#  & Limited & .NET 8.0.117 & ASP.NET Core & MsQuic \\
  TypeScript & \XSolidBrush  & Node.js v20.16.0 & Express & Ajv validator\\
  Ruby & Limited & CRuby 3.2.0 & Sinatra & FastImage \\
  R & \XSolidBrush & R 4.3.3 & RestRserve & ggplot2 \\
  Scala  & \XSolidBrush & Scala 3.5.0 & Scalatra & GeoTrellis \\
  Kotlin  & \XSolidBrush & Kotlin 2.0.10 & Javalin & Kotlin serialization \\
 \bottomrule
\end{tabular}
\begin{tablenotes}
    \footnotesize
    \item[*] Ranked by \textit{The RedMonk Programming Language Rankings}~\cite{redmonk}.
\end{tablenotes}
}
\end{threeparttable}
\label{tab:languages}
\end{table}

%% file: tables/bm_intro.tex
\begin{table}[t!]\small
\centering
\caption{\add{Task overview in \textit{The Computer Language Benchmarks Game}~\cite{benchmarksgame}.}}
\vspace{-5pt}
\add{
\begin{tabular}{ll}
 \toprule
 \textbf{Task} & \textbf{Description} \\
 \midrule
  \texttt{binary-trees} & Allocate and deallocate binary trees \\ 
  \texttt{fannkuch-redux} & Indexed-access to tiny integer-sequence \\ 
  \texttt{fasta} & Generate and write random DNA sequences \\ 
  \texttt{k-nucleotide} & Hashtable update and k-nucleotide strings \\ 
  \texttt{mandelbrot} & Generate Mandelbrot set portable bitmap file \\ 
  \texttt{n-body} & Double-precision N-body simulation \\
  \texttt{pidigits} & Streaming arbitrary-precision arithmetic \\
  \texttt{regex-redux} & Match DNA 8-mers and substitute magic patterns \\
  \texttt{reverse-complement} & Read DNA sequences - write their reverse-complement \\
  \texttt{spectral-norm} & Eigenvalue using the power method \\
 \bottomrule
\end{tabular}
}
\label{tab:bm_intro}
\end{table}

%% file: tables/bm_pass.tex
\begin{table}[t!]\small
\centering
\caption{\add{Benchmark pass rates on \tool across 10 managed languages.}}
\vspace{-5pt}
\add{
\begin{tabular}{l|cccc|c}
 \toprule
 \textbf{Language} & \textbf{\#Runnable} & \textbf{\#Passed} & \textbf{\#Failed} & \textbf{Pass Rate} & \textbf{Failure Mode}\\
 \midrule
  JavaScript & 23 & 23 & 0 & 100\% & - \\
  Python     & 30 & 27 & 3 & 90\% & Runtime error \\
  Java       & 48 & 48 & 0 & 100\% & - \\
  PHP        & 31 & 22 & 9 & 70.97\% & Output mismatch \\
  C\#        & 41 & 41 & 0 & 100\% & - \\
  TypeScript & 10 & 10 & 0 & 100\% & - \\
  Ruby       & 31 & 31 & 0 & 100\% & - \\
  R          & 40 & 40 & 0 & 100\% & - \\
  Scala      & 14 & 14 & 0 & 100\% & - \\
  Kotlin     & 10 & 10 & 0 & 100\% & - \\
  \midrule
  \textbf{Total} & 278 & 266 & 12 & 95.68\% & \\
 \bottomrule
\end{tabular}
}
\label{tab:bm_pass}
\end{table}

%% file: figures/python_fail_exp.tex
\definecolor{pybg}{RGB}{248,248,248}
\definecolor{pyframe}{RGB}{220,220,220}
\definecolor{pykw}{RGB}{0,92,197}      
\definecolor{pybuiltin}{RGB}{126,63,152}
\definecolor{pydecor}{RGB}{170,0,0}    
\definecolor{pystr}{RGB}{0,128,0}      
\definecolor{pycmt}{RGB}{120,120,120}  
\definecolor{pynum}{RGB}{176,80,0}     
\definecolor{pyop}{RGB}{90,90,90}      

\lstdefinestyle{compactpython}{
  language=Python,
  basicstyle=\ttfamily\scriptsize,
  columns=fullflexible,
  keepspaces=true,
  upquote=true,
  frame=single,
  rulecolor=\color{pyframe},
  framerule=0.3pt,
  backgroundcolor=\color{pybg},
  xleftmargin=2pt,
  xrightmargin=2pt,
  aboveskip=2pt,
  belowskip=2pt,
  numbers=none,
  showstringspaces=false,
  breaklines=true,
  breakatwhitespace=true,
  keywordstyle=\color{pykw}\bfseries,
  commentstyle=\color{pycmt}\itshape,
  stringstyle=\color{pystr},
  alsoletter={_}, 
  morekeywords={True,False,None,self,cls},
  morekeywords=[2]{print,len,range,map,filter,zip,enumerate,int,float,str,bool,list,dict,set,tuple,  sum,min,max,sorted,open,type,isinstance,super,any,all},
  keywordstyle=[2]\color{pybuiltin},
  morekeywords=[3]{@app.route,@app.get,@app.post,@app.put,@app.delete,@staticmethod,@classmethod,@property},
  keywordstyle=[3]\color{pydecor}\bfseries,
  emph={:,=,(,),[,],\{,\},.,->},
  emphstyle=\color{pyop},
}

\begin{figure}[t]
  \centering
  \begin{subfigure}[b]{0.48\linewidth}
    \begin{minipage}[b]{\linewidth}
\begin{lstlisting}[style=compactpython]
from multiprocessing import Pool
...
def multiply_AtAv(u):
    r = range(len(u))
    tmp = pool.starmap(A_sum, zip(repeat(u), r))
    return pool.starmap(At_sum, zip(repeat(tmp), r))
...
if __name__ == "__main__":
    with Pool(processes=4) as pool:
        main(int(sys.argv[1]))
\end{lstlisting}
    \end{minipage}
    \caption{Benchmark snippet (\texttt{spectralnorm\_py\_4.py})}\label{fig:python_fail_bm}
  \end{subfigure}  
  \hfill 
  \begin{subfigure}[b]{0.48\linewidth}
    \begin{minipage}[b]{\linewidth}
\begin{lstlisting}[style=compactpython]
# Flask service routing logic
@app.route("/spectralnorm_py_4")
def call_spectralnorm_py_4():
    from python_benchmark.spectralnorm_py_4 import main
    from multiprocessing import Pool
    ...
    with Pool(processes=4) as pool:
        main(int(request.args.get("a")))
    return "over\n"
\end{lstlisting}
    \end{minipage}
    \caption{Service wrapper snippet (\texttt{app.py})}\label{fig:python_fail_app}
  \end{subfigure}

  \caption{\add{A representative Python failed case. The benchmark assumes a module-level \texttt{pool} initialized only under \texttt{\_\_main\_\_}, while the service wrapper creates a local \texttt{Pool} not visible to the imported module.}
  }
  \label{fig:python_fail_exp}
\end{figure}

%% file: figures/php_fail_exp.tex
\lstdefinestyle{compactphp}{
  language=PHP,
  basicstyle=\ttfamily\scriptsize,
  columns=fullflexible,
  keepspaces=true,
  upquote=true,
  frame=single,
  rulecolor=\color{pyframe},
  framerule=0.3pt,
  backgroundcolor=\color{pybg},
  xleftmargin=2pt,
  xrightmargin=2pt,
  aboveskip=2pt,
  belowskip=2pt,
  numbers=none,
  showstringspaces=false,
  breaklines=true,
  breakatwhitespace=true,
  keywordstyle=\color{pykw}\bfseries,
  commentstyle=\color{pycmt}\itshape,
  stringstyle=\color{pystr},
  morekeywords={
    declare,strict_types,namespace,use,as,
    class,trait,interface,extends,implements,
    public,protected,private,static,final,abstract,readonly,
    function,fn,return,yield,throw,try,catch,finally,
    if,elseif,else,switch,case,default,
    for,foreach,while,do,break,continue,
    match,new,clone,this,parent,self,
    global,var,const,unset,isset,empty,
    true,false,null,
    echo,print,include,include_once,require,require_once
  },
  morekeywords=[2]{
    array_merge,array_map,array_filter,array_reduce,count,in_array,
    explode,implode,strpos,substr,strlen,trim,strcmp,strtolower,strtoupper,
    json_encode,json_decode,
    preg_match,preg_replace,
    file_get_contents,file_put_contents,fopen,fclose,
    getenv,putenv,
    date,time,
    header,http_response_code
  },
  keywordstyle=[2]\color{pybuiltin},
}

\begin{figure}[t]
  \centering
  \begin{subfigure}[b]{0.48\linewidth}
  \begin{minipage}[b]{\linewidth}
\begin{lstlisting}[style=compactphp]
function startWorker(...) {
    $pid = pcntl_fork();
    ...
}
...
function waitWorkers(&$PIDs) {
    foreach ($PIDs as $PID) {
        $pid = pcntl_waitpid($PID, $status);
    }
}
\end{lstlisting}
    \caption{Benchmark snippet (\texttt{binarytrees\_php\_6.php})}\label{fig:php_fail_a}
  \end{minipage}
  \end{subfigure}
  \hfill
  \begin{subfigure}[b]{0.48\textwidth}
  \begin{minipage}[b]{\linewidth}
\begin{lstlisting}[style=compactphp]
$app->get('/binarytrees_php_6', function (...) {
    ...
    $workersPIDs = [];
    foreach ($depthIterations as $depth => $iter) {
        startWorker($depth, $iter, $workersPIDs,...);
    }
    waitWorkers($workersPIDs);
    ...
    return $response;
});
\end{lstlisting}
\caption{Service wrapper snippet (\texttt{server.php})}\label{fig:php_fail_b}
  \end{minipage}
  \end{subfigure}

  \caption{\add{A representative PHP failed case. The benchmark relies on OS-level process management (\eg \texttt{pcntl\_fork}/\texttt{pcntl\_waitpid}) to synchronize worker processes, which may be restricted in web services.}}
  \label{fig:php_fail}
\end{figure}

%% file: tables/exec_time.tex
\begin{table}[t!]\small
\centering
\caption{Execution time statistics of Python benchmarks on \base and \tool (in seconds).}
\vspace{-5pt}
\begin{threeparttable}
\resizebox{\linewidth}{!}{
\begin{tabular}{lccc|lccc|lccc}
 \toprule
 \multicolumn{4}{c}{\textbf{Compute-intensive Tasks}} & \multicolumn{4}{c}{\textbf{I/O-intensive Tasks}} & \multicolumn{4}{c}{\textbf{Memory-intensive Tasks}} \\
 \midrule
 \textbf{BM} & PY-WE\tnote{*} & \tool & $\uparrow\times$ & 
 \textbf{BM} & PY-WE & \tool & $\uparrow\times$ & 
 \textbf{BM} & PY-WE & \tool & $\uparrow\times$ \\
 \midrule 
 
 pystone & 96.551 & 0.129 & 748 &
 file & 0.241 & 0.002 & 120 &
 alloc & 66.194 & 0.184 & 359 \\
 
 nqueens & 765.146 & 1.434 & 553 &
 json & 13.325 & 0.026 & 512 &
 string & 1.207 & 0.002 & 603 \\

 prime & 8.332 & 0.010 & 833 &
 text & 22.534 & 0.050 & 450 &
 large & 39.322 & 0.080 & 491 \\

 matrix & 70.985 & 0.074 & 959 &
 list & 84.755 & 0.153 & 553 &
 object & 118.213 & 0.319 & 370 \\

 fibonacci & 1377.160 & 1.934 & 712 &
 dict & 16.701 & 0.043 & 388 &
 recur & 1.068 & 0.002 & 534 \\

 \midrule
 \textbf{Avg.} & 463.635 & 0.716 & 647 &
 \textbf{Avg.} & 27.511 & 0.055 & 500 &
 \textbf{Avg.} & 45.201 & 0.117 & 386 \\
 
 \bottomrule
\end{tabular}
}
\begin{tablenotes}
    \footnotesize{\item[*] PY-WE refers to \base.}
\end{tablenotes}
\end{threeparttable}
\label{tab:exec_time}
\vspace{-5pt}
\end{table}

%% file: tables/exec_py_apps.tex
\begin{table}[t!]\small
\centering
\caption{\add{Execution time statistics of real-world Python libraries (in seconds).}}
\vspace{-5pt}
\begin{threeparttable}
\add{
\begin{tabular}{cccc|cccc}
 \toprule
 \multicolumn{4}{c}{\textbf{jsonschema ($n{=}1024$)}} & \multicolumn{4}{c}{\textbf{markdown ($n{=}256$)}} \\
 \midrule
 \textbf{Chunk ($k$)} & PY-WE & \tool & $\uparrow\times$ & 
 \textbf{Chunk ($k$)} & PY-WE & \tool & $\uparrow\times$ \\
 \midrule 
 1 & 63.251 & 0.097 & 652 &
 1 & 99.566 & 0.225 & 443
 \\
 8 & 63.100 & 0.090 & 701 &
 8 & 98.929 & 0.218 & 454
 \\
 32 & 63.840 & 0.089 & 717 &
 32 & 103.344 & 0.228 & 453
 \\
 \midrule
 \textbf{Avg.} & 63.397 & 0.092 & 689 &
 \textbf{Avg.} & 100.613 & 0.224 & 449 \\
 \bottomrule
\end{tabular}
\begin{tablenotes}
\footnotesize
\item $n$: total inputs (docs for \texttt{jsonschema}, texts for \texttt{markdown}). 
\item $k$: number of chunks/invocations (total input size $n$ is fixed).
\end{tablenotes}
}
\end{threeparttable}
\label{tab:exec_py_apps}
\vspace{-5pt}
\end{table}

%% file: tables/overhead.tex
\begin{table}[t!]\small
\centering
\caption{Communication overhead statistics across Python benchmarks (in milliseconds).}
\vspace{-5pt}
\begin{threeparttable}
\resizebox{\linewidth}{!}{
\begin{tabular}{l|cc|cccc}
 \toprule
 \textbf{BM} &  $\mathbf{T_{total}}$ & $\mathbf{T_{exec}}$ & $\mathbf{O_{client}}$ & $\mathbf{O_{network}}$ & $\mathbf{O_{server}}$ & $\mathbf{O_{total}}$ \\
 
 \midrule
 
 pystone & 131.725 (100\%) & 129.706 (98.47\%) & 0.134 (0.10\%) & 1.794 (1.36\%) & 0.091 (0.07\%) & \textbf{2.019 (1.53\%)} \\

 nqueens & 1442.345 (100\%) & 1440.165 (99.85\%) & 0.235 (0.02\%) & 1.837 (0.13\%) & 0.108 (0.01\%) & \textbf{2.180 (0.15\%)} \\

 prime & 13.553 (100\%) & 11.434 (84.37\%) & 0.135 (1.00\%) & 1.870 (13.80\%) & 0.114 (0.84\%) & \textbf{2.119 (15.63\%)} \\

 matrix & 78.456 (100\%) & 76.341 (97.30\%) & 0.146 (0.19\%) & 1.852 (2.36\%) & 0.117 (0.15\%) & \textbf{2.115 (2.70\%)} \\

 fibonacci & 1905.187 (100\%) & 1902.936 (99.88\%) & 0.140 (0.01\%) & 2.013 (0.11\%) & 0.108 (0.01\%) & \textbf{2.261 (0.12\%)} \\

 \midrule

 \textbf{Avg.} & 714.253 (100\%) & 712.116 (99.70\%) & 0.158 (0.02\%) & 1.873 (0.26\%) & 0.108 (0.02\%) & \textbf{2.139 (0.30\%)} \\
 
 \bottomrule
\end{tabular}
}
\begin{tablenotes}
    \footnotesize
    \item 
    $\mathbf{T_{total}}{=}T(\text{\ding{192}}{+}\text{\ding{193}})$, 
    $\mathbf{T_{exec}}{=}T(\text{\ding{196}})$,
    $\mathbf{O_{client}}{=}T(\text{\ding{192}})$,
    $\mathbf{O_{network}}{=}T(\text{\ding{194}}{+}\text{\ding{198}})$, 
    $\mathbf{O_{server}}{=}T(\text{\ding{195}}{+}\text{\ding{197}})$,
    $\mathbf{O_{total}}{=}\mathbf{O_{client}}{+}\mathbf{O_{network}}{+}\mathbf{O_{server}}$.
\end{tablenotes}
\end{threeparttable}
\label{tab:overhead}
\end{table}

%% file: tables/overhead_py_apps.tex
\begin{table}[t!]\small
\centering
\caption{\add{Communication overhead statistics of real-world Python libraries (in milliseconds).}}
\vspace{-5pt}
\resizebox{0.9\linewidth}{!}{
\add{
\begin{tabular}{cccc|cccc}
 \toprule
 \textbf{jsonschema} &  $\mathbf{T_{total}}$ & $\mathbf{T_{exec}}$ & $\mathbf{O_{total}}$ & 
 \textbf{markdown} &  $\mathbf{T_{total}}$ & $\mathbf{T_{exec}}$ & $\mathbf{O_{total}}$\\
 \midrule
 $k{=}1$ & 103.846 & 91.986 & \textbf{11.860 (11.42\%)} &
 $k{=}1$ & 226.736 & 221.805 & \textbf{4.931 (2.17\%)} \\
 $k{=}8$ & 101.871 & 90.767 & \textbf{11.104 (10.90\%)} &
 $k{=}8$ & 223.549 & 218.172 & \textbf{5.377 (2.41\%)} \\
 $k{=}32$ & 97.963 & 86.325 & \textbf{11.638 (11.88\%)} &
 $k{=}32$ & 222.045 & 217.217 & \textbf{4.828 (2.17\%)} \\
 \midrule
 \textbf{Avg.} & 101.227 & 89.693 & \textbf{11.534 (11.39\%)} & 
 \textbf{Avg.} & 224.110 & 219.065 & \textbf{5.045 (2.25\%)} \\
 \bottomrule
\end{tabular}
}
}
\label{tab:overhead_py_apps}
\vspace{-5pt}
\end{table}

%% file: tex/5-dis.tex
\section{Discussion}\label{sec:dis}

\subsection{\add{Limitations}}

\add{\tool currently supports invoking external libraries initiated from Wasm programs, but it does not support callbacks from external runtimes back into Wasm. This limitation arises from our design goal of enabling a simple and extensible external library linking mechanism, in which Wasm programs act as clients that invoke external libraries using a request-response interaction pattern. Supporting callbacks would require a more tightly coupled and bidirectional execution model across runtimes, which is beyond the scope of the current design of \tool. Consequently, \tool is best suited for scenarios with clearly defined library boundaries, while integration patterns that rely heavily on callbacks or event-driven interactions are not directly supported.}

\add{\tool assumes that external libraries execute in trusted managed language runtimes outside the Wasm sandbox. While the Wasm-side orchestration logic remains sandboxed, the execution of external libraries relies on the security mechanisms provided by the host environment. This design shifts the security boundary compared to approaches that compile managed runtimes entirely into Wasm and therefore inherit Wasm’s sandbox guarantees. As a result, \tool is most appropriate for controlled deployment settings, such as internal services or trusted library integration, rather than for scenarios involving untrusted third-party code.}

\add{Although \tool avoids the complexity of runtime nesting and execution model extensions, it introduces additional usability challenges at the developer level. Developers must reason about cross-language interfaces, parameter marshaling, and error propagation across different runtimes, which can increase cognitive load and complicate debugging. 
To mitigate these challenges, \tool is designed to minimize client-side development effort by providing unified invocation interfaces, structured metadata, and automated cross-runtime communication. Nevertheless, cross-language integration inherently requires additional coordination across execution environments, and this usability trade-off may still affect development experience in complex integration scenarios.}

\subsection{Threats to Validity}

The first threat relates to the generalizability of \tool's external library linking design, which may not accommodate all types of application scenarios. To maximize generality, we adopted standard HTTP RESTful interfaces to ensure protocol compatibility and utilized JSON as the primary data serialization format for its widespread support across programming ecosystems. While this design may not be optimal for all use cases, it provides a robust foundation for the majority of cross-language invocation scenarios encountered in practice.

The second threat concerns the potential limitations in language extensibility evaluation due to constrained test case coverage. To mitigate this threat, we based our language selection on the RedMonk rankings to include mainstream programming languages that represent different computational paradigms. Our evaluation encompassed diverse runtime characteristics, ensuring that our findings reflect the extensibility of \tool across a wide spectrum of managed language environments, though certain niche or emerging languages may require additional validation.

The third threat involves potential biases in performance evaluation, including baseline selection and benchmark construction. We mitigated this concern by choosing the representative \base runtime as our baseline to ensure credibility. 
Due to functional limitations of the baseline environment, we constructed a dataset that maintains functional equivalence while ensuring executability in both test environments. Our dataset incorporates diverse workload types and employs statistical methods to ensure the reliability of our performance comparisons.

%% file: tex/6-rw.tex
\section{Related Work}\label{sec:rw}

\subsubsection*{\textbf{Wasm Runtime Features}}
Wasm has attracted a lot of research interest as a promising technique since it was initially introduced~\cite{stievenart2022static,he2023eunomia,lehmann2023sa}.
It has been widely applied on both the web side~\cite{reiser2017accelerate,jangda2019not,romano2022wobfuscator} and the server side~\cite{mendki2020evaluating,nurul2021nomad,gackstatter2022pushing} in recent years.
To make Wasm better adapted to various application scenarios, many studies aim at analyzing and improving Wasm runtime features, including safety~\cite{lehmann2020everything,geller2024indexed,bosamiya2022provably,johnson2023wave}, efficiency~\cite{romano2023function,jiang2023revealing,liu2023exploring,jiang2025distinguishability,zeng2026debugging}, lightweight~\cite{menetrey2021twine,sebrechts2022adapting,nakakaze2022retrofitting}, etc.

Among existing research on Wasm runtime features, safety is the most compelling aspect~\cite{zhang2024research}.
Lehmann \etal~\cite{lehmann2020everything} first studied the vulnerabilities in Wasm binaries, and provided a set of vulnerable applications along with end-to-end exploits.
Narayan \etal designed Swivel~\cite{narayan2021swivel}, a new compiler framework for hardening Wasm against Spectre attacks.
\delete{Other Wasm runtimes aimed at enhancing the safety feature include Acctee~\cite{goltzsche2019acctee}, PKUWA~\cite{lei2023put}, CWASI~\cite{marcelino2023cwasi}, etc.}
Runtime efficiency is another key feature widely studied for Wasm.
Titzer introduced a fast in-place Wasm interpreter ~\cite{titzer2022fast}, which improves the runtime efficiency by compiling Wasm to machine code without rewriting or a separate format.
Moron \etal~\cite{moron2023support} presented a microcontroller-compatible Wasm runtime that supports JIT compilation to improve the execution speed.
In addition, lightweight Wasm runtimes are also an attractive direction.
For example, Sledge~\cite{gadepalli2020sledge} is a lightweight Wasm-based serverless framework optimized for low startup time, bursty client request rates, and short-lived computations.

As Wasm is a compilation target for programming languages, the language support capability is also a key feature to consider.
However, Wasm's current language support is still immature, especially for managed languages~\cite{hilbig2021empirical}.

\subsubsection*{\add{\textbf{Managed Language Support in Wasm}}}
\add{Supporting managed languages in Wasm remains challenging due to their reliance on complex runtime systems. A common line of work adopts runtime nesting, in which the language runtime is first compiled to Wasm, allowing source programs to execute on that Wasm-based runtime. Representative examples include Python WASI support~\cite{py-wasmedge}, and lightweight JavaScript engines such as QuickJS~\cite{quickjs}. While runtime nesting enables basic execution of managed languages within Wasm environments, prior studies have observed limitations in extensibility, performance, and compatibility with language features.}

\add{More recently, the Wasm community has explored extending the execution model itself to better accommodate managed languages. The WebAssembly Garbage Collection (WasmGC) proposal~\cite{wasmgc} introduces better support for garbage-collected objects and typed references, enabling managed languages to be compiled more directly to Wasm without embedding an entire language runtime. 
However, WasmGC-based approaches rely on dedicated compiler support and GC-enabled Wasm engines, and their applicability depends on the availability and maturity of such runtimes in target deployment environments~\cite{steiner2024toward,guellil2025optimizing}. Such reliance may limit its applicability in environments where modifying or upgrading the Wasm execution stack is undesirable.
}

\subsubsection*{\textbf{Dynamic Linking}}
\add{Beyond extending the Wasm execution model, dynamic linking~\cite{franz1997dynamic,ho1991approach} is another alternative mechanism.} It is a technique for linking software libraries to target programs at execution time, offering benefits such as code reuse, library updates without source modification, and enhanced security through address space randomization~\cite{shacham2004effectiveness}.
Many existing studies adopt dynamic linking techniques to optimize software systems~\cite{de2005link,altinay2020binrec,dunkels2006run,dong2009dynamic}.
For example, Dunkels \etal~\cite{dunkels2006run} implemented an in-situ run-time dynamic linker and loader for reprogramming resource-constrained wireless sensor nodes. 
Dong \etal~\cite{dong2009dynamic} further proposed a holistic dynamic linking and loading mechanism in networked embedded systems for minimal code size, efficient runtime speed, and kernel-user isolation.
There are also extensive studies devoted to providing better algorithms and architectures for dynamic linking~\cite{agrawal2015architectural,malabarba2000runtime,ren2022dynamic,bartell2020guided}.

In the Wasm community, dynamic linking has been applied for function extension and performance optimization~\cite{makitalo2021bringing,makitalo2021webassembly,lehmann2019wasabi}.
Wasm programs are organized into modules, which can be dynamically linked to create modular applications.
Mäkitalo \etal~\cite{makitalo2021bringing} presented a dynamic linking system for Wasm modules and studied its performance.
Wen \etal introduced WasmSlim~\cite{wen2023wasmslim}, which transforms Wasm into a slim main module and dynamically linked secondary modules to reduce binary size and improve startup speed.
\delete{In this work, we extend managed language support for Wasm by dynamically linking external libraries.}

\subsubsection*{\add{\textbf{HTTP-Based Cross-Language Invocation}}}

\add{A substantial body of prior work adopts HTTP-based invocation as a practical mechanism for cross-language integration~\cite{slee2007thrift,grimmer2018cross,li2016design}. Widely used frameworks such as FastAPI for Python~\cite{fastapi} and Spring Boot for Java~\cite{springboot} enable language-specific libraries and services to be exposed through RESTful APIs, allowing clients written in other languages to invoke them via standard HTTP interfaces. 
Beyond individual frameworks, HTTP and RPC-based cross-language invocation has been extensively studied in the context of service-oriented architectures and microservices~\cite{papazoglou2007service,dragoni2017microservices,newman2021building}, where language-agnostic communication protocols serve as the foundation for polyglot systems. These approaches typically assume independently deployed services and focus on coarse-grained service invocation across language boundaries.}

\add{In the context of Wasm, this work explores leveraging HTTP-based invocation to integrate managed-language libraries with Wasm programs. 
In this setting, HTTP serves not merely as a general-purpose service interface, but as an enabling mechanism for dynamically linking external libraries to Wasm modules at runtime. 
This positioning distinguishes HTTP-based invocation in the Wasm setting from its traditional role in microservice architectures and motivates its use as a lightweight alternative to runtime nesting or execution-model extensions.}

%% file: tex/7-con.tex
\section{Conclusion}\label{sec:con}

This paper presents \tool, a novel framework that addresses the critical challenge of managed language support in Wasm through external library linking. Unlike existing runtime nesting approaches that suffer from limited language compatibility and performance overhead, \tool introduces a client-server architecture that \delete{maintains the security guarantees of Wasm} \add{keeps the Wasm module sandboxed} while enabling \delete{seamless} \add{efficient} integration with diverse language runtimes\delete{, and provides practical LLM-assisted development tools for cross-language programming}. Our evaluation demonstrates that \tool successfully supports ten managed languages without framework modifications, achieves near-native execution performance with low communication overhead.
This work establishes a foundation for multi-language edge computing and opens new directions for performance optimization and cloud-native deployment in Wasm ecosystems.

\section*{Data Availability}
Our source code and experiment data are available at \url{https://figshare.com/s/5da0d15030538b69fcca}.